\documentclass[%
 reprint,
superscriptaddress,
longbibliography,
 aip,
 amsmath,amssymb,
floatfix,
]{revtex4-2}
\setcitestyle{numbers,square}

\usepackage{braket}
\usepackage{graphicx}
\usepackage{dcolumn}
\usepackage{bm}
\usepackage[normalem]{ulem} 
\usepackage{color}

\usepackage{hyperref}
\hypersetup{
    colorlinks,%
    citecolor=blue,%
    linkcolor=blue,%
    urlcolor=blue
}

\newcommand{\blue}[1]{{#1}}

\begin{document}
\title{Unveiling the dopant segregation effect at hematite interfaces}

\author{Felipe C. de Lima}
\email{felipe.lima@lnnano.cnpem.br}
\affiliation{Brazilian Nanotechnology National Laboratory, \\ C.P. 6192, 13083-970 Campinas, SP, Brazil}

\author{Gabriel R. Schleder}
\affiliation{Brazilian Nanotechnology National Laboratory, \\ C.P. 6192, 13083-970 Campinas, SP, Brazil}
\affiliation{Federal University of ABC, Santo Andr\'e, 09210-580, SP, Brazil}

\author{Jo\~ao B. Souza Junior}
\affiliation{Brazilian Nanotechnology National Laboratory, \\ C.P. 6192, 13083-970 Campinas, SP, Brazil}

\author{Flavio L. Souza}
\affiliation{Brazilian Nanotechnology National Laboratory, \\ C.P. 6192, 13083-970 Campinas, SP, Brazil}
\affiliation{Federal University of ABC, Santo Andr\'e, 09210-580, SP, Brazil}

\author{Fabrício B. Destro}
\affiliation{Federal University of São Carlos (UFSCar), Graduate Program in Materials Science and Engineering, Department of Materials Engineering, 13565-905, São Carlos, SP, Brazil}

\author{Roberto H. Miwa}
\affiliation{Instituto de F\'isica, Universidade Federal de Uberl\^andia, \\ C.P. 593, 38400-902, Uberl\^andia, MG,  Brazil}

\author{Edson R. Leite}
\affiliation{Brazilian Nanotechnology National Laboratory, \\ C.P. 6192, 13083-970 Campinas, SP, Brazil}
\affiliation{Departamento de Química, Universidade Federal de São Carlos, 13565-905, Brazil}

\author{Adalberto Fazzio}
\affiliation{Brazilian Nanotechnology National Laboratory, \\ C.P. 6192, 13083-970 Campinas, SP, Brazil}
\affiliation{Federal University of ABC, Santo Andr\'e, 09210-580, SP, Brazil}

\date{\today}

\begin{abstract}

Understanding the effects of atomic structure modification in hematite photoanodes is essential for the rational design of high-efficiency functionalizations. Recently it was found that interface modification with Sn/Sb segregates considerably increases hematite photocatalytic efficiency. However, the understanding of the different electronic effects of these modifications at the atomic level is still lacking. This letter describes the segregation effects of two different dopants—Sn and Sb—on both the solid-solid (grain-boundaries) and solid-liquid interfaces (surfaces) of hematite. Within an \textit{ab initio} approach, we quantitatively extract the potential barrier reduction on polycrystalline interfaces due to the dopant, which causes an increase in the inter-grain electron transport. Concomitantly, the dopants' segregation on hematite surfaces results in a decrease of the oxygen vacancy formation energy. Such vacancies lead to the experimentally observed rise of the flat-band potential. The comprehension of the electronic effects of dopants on both types of interfaces explains the experimental peak efficiency of interface-modified hematite with dopant segregates, also enabling the control and design of interfaces for different higher-efficiency applications.

\end{abstract}

\maketitle

Hematite ($\alpha$-Fe2O3), an n-type semiconductor with high photoelectrochemical stability, is considered the most promising material to be used as a photoanode for water splitting (with a potential to convert $16.8\%$ of the sun’s energy into hydrogen) \cite{CSCsivula2011}. However, problems related to low electronic conductivity and poor kinetics of charge transfer (mainly holes) at the semiconductor-liquid interface causes an efficiency in the water oxidation process far below the theoretical prediction \cite{CRwalter2010}. An alternative found to overcome these problems on hematite is the insertion of metallic and semi-metallic elements, such as Sn and Sb \cite{EESgoncalves2014, AFMhufnagel2018, ACSAMIannamali2018, JMCAnogueira2019}.

\begin{figure}[!hb]
\includegraphics[width=0.82\columnwidth]{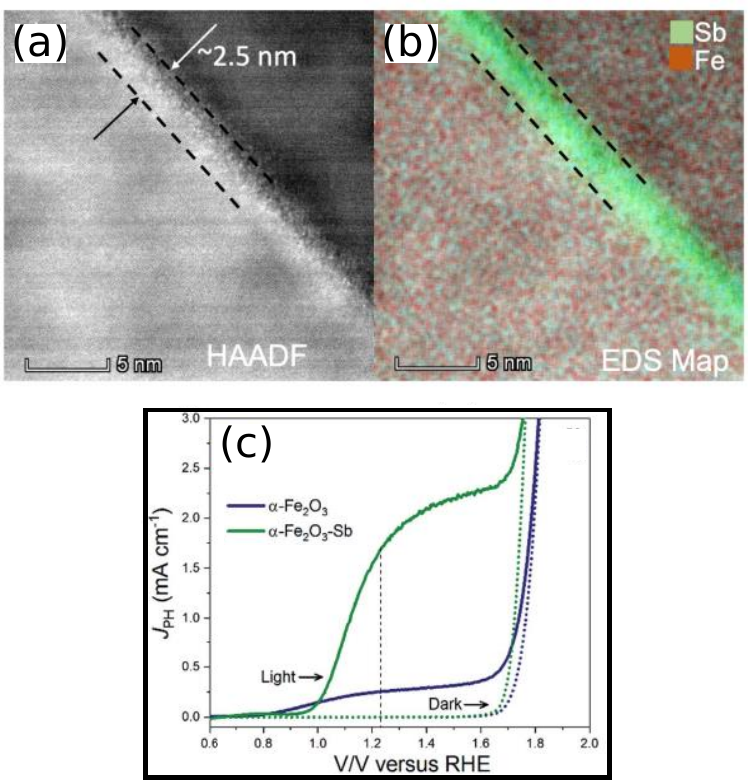}
\caption{\label{fig:exp} (a) HAADF image of a grain boundary region of a Sb-doped hematite system. The bright contrast at grain boundary area is consistent with the Sn segregation; (b) STEM/EDS image showing an EDS elemental map distribution. This analysis confirms the Sb segregation at the grain boundary area. The STEM/EDS analysis was performed in a ceramic polycrystalline hematite doped with $2\%$ by weight of Sb$_2$O$_3$ and sintered at $1300^{o}$C during 2 h; (c) photoelectrochemical (J-V curve) measurement of hematite film doped with Sb (adapted from \cite{JMCAnogueira2019} with permission of The Royal Society of Chemistry).}
\end{figure}

\begin{figure*}
\includegraphics[width=1.43\columnwidth]{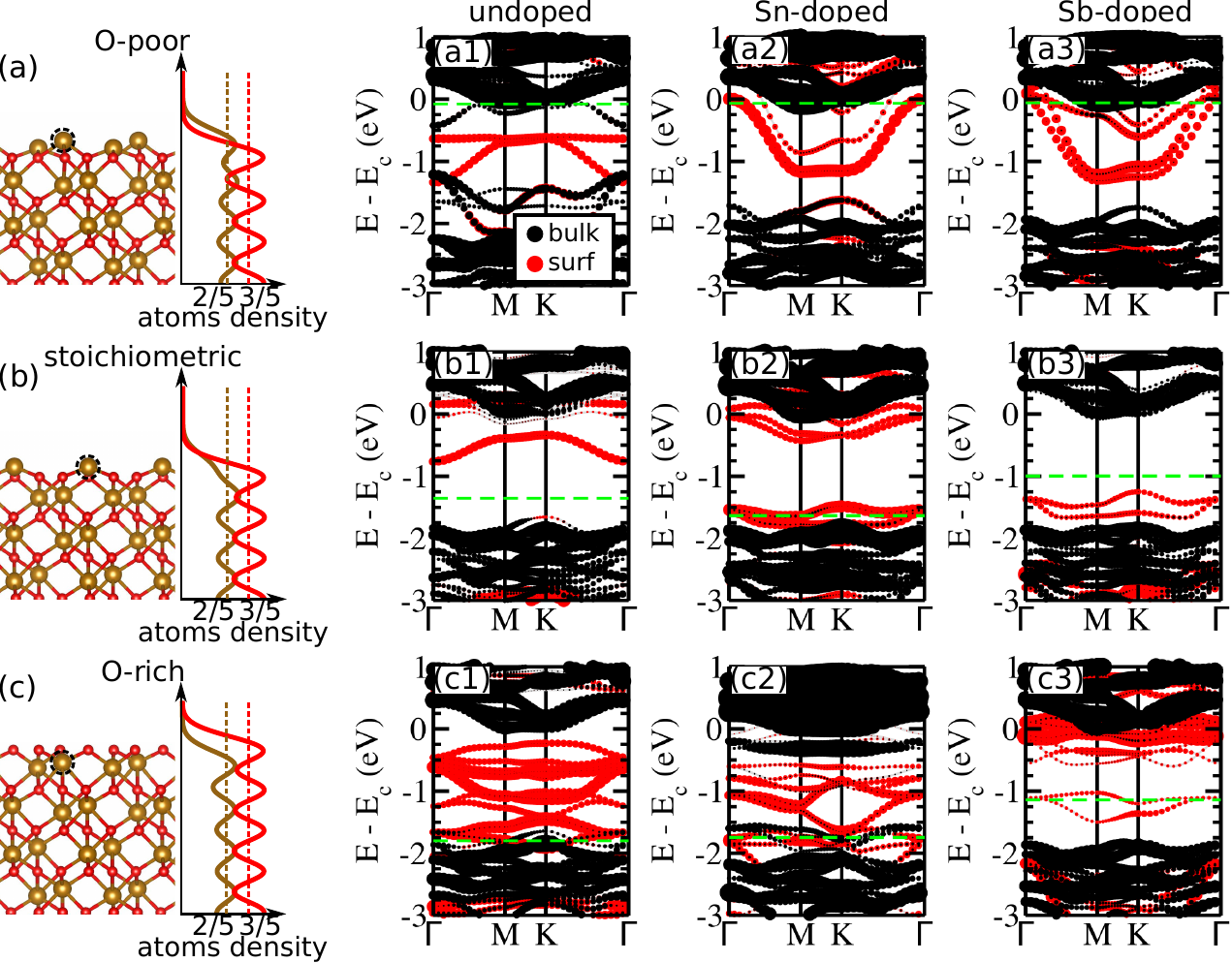}
\caption{\label{fig:surf-band-001} Slab atomic structure with (a) O-poor, (b) stoichiometric and (c) O-rich terminations, with O (Fe) atoms shown in red (gold). The right panels show the atomic density, $\eta$, for the oxygen (red) and Fe atoms (gold). Band structure of (001) hematite slabs in three surface conditions: (a1)-(c1) Undoped surface, (a2)-(c2) Sn-doped surface, (a3)-(c3) Sb-doped surface. The black and red circles display the contribution of the bulk and surface atoms to each state ($|\langle \phi_{\rm atoms} | \Psi_{n,\,{\bm k}} \rangle|^2$). We define the bottom of the conduction band ($E_c$) as a reference level, and the green dashed lines pinpoint the Fermi level.}
\end{figure*}

The ionic nature of hematite -- which is associated with Schottky intrinsic ionic disorder -- leads to dopants segregation at grain boundaries and free surfaces \cite{PCCPsoares2016, PCCPengel2014}. Consequently, the physical-chemistry structure of the solid-solid (grain boundary in a polycrystalline film) and solid-liquid interfaces is modified. Recently, the addition of Sn on hematite was shown to preferentially segregate at the grain boundary region \cite{AEMsoares2019}. Similarly, sintered hematites with $2\%$ by weight of Sb$_2$O$_3$ also leads to Sb segregation at the grain boundary region \cite{JMCAnogueira2019}. The scanning transmission electron microscopy (STEM) with high angle annular dark-field (HAADF) image [Fig. ~\ref{fig:exp}(a)], as well as the energy-dispersive spectroscopy [EDS Map, Fig. ~\ref{fig:exp}(b)] show a Sb layer of approximately $2.5$\ nm thick. Both dopants segregating at grain boundary region shows a reduction in total electronic resistivity, with a specific decrease in the grain boundary resistance \cite{AEMsoares2019, destro_2020}. Sn or Sb promotes a decrease of 3 to 5 orders of magnitude in the electronic resistivity on hematite \cite{AEMsoares2019, destro_2020}. The dopant segregation also modifies the solid-liquid interfaces, changing the photoelectrochemical properties of hematite photoanodes \cite{EESgoncalves2014, JMCAnogueira2019}. Indeed, as illustrated in Fig.~\ref{fig:exp}(c), the addition of Sb promotes a significant increase in photocurrent ($J_{\rm PH}$). Moreover, an increase in the flat band potential ($V_{\rm fb}$) is also reported, suggesting the formation of surface states \cite{JMCAnogueira2019}. The Sn doped hematite shows similar results, i.e., an increase in $J_{\rm PH}$ and $V_{\rm fb}$ \cite{EESgoncalves2014}.

Despite efforts dedicated to understanding the role of dopants from macroscopic to atomic-scale analysis --- revealing the atomic scale segregation at the hematite interfaces leading to those discussed effects --- a profound lack of information remains regarding its electronic structure. This letter is devoted to unravelling how the electronic structure is affected by dopant segregation at hematite interfaces --- assigned by decreasing the grain boundary resistance, with the concomitant creation of surface states at the solid-liquid interface. By employing density functional theory (DFT) calculations, we describe the electronic effects of Sb/Sn-doped hematite surfaces and grain boundaries. Here we have shown that the favorable conditions for O-vacancy formation in the doped surfaces rule the flat-band potential increase. By tracking the effective electronic potential across hematite grain-boundaries, we predict the decrease of the effective potential barrier height, increasing the electron transport rate between grains.

{\it Methods.} DFT calculations \cite{dft1964,dft1965,review} are performed using the Viena Ab-initio Simulation Package (VASP) implementation \cite{CMSkresse1996}. The exchange-correlation term was described using the GGA functional proposed by Perdew, Burke, and Ernzerhof (PBE) \cite{PRLperdew1996, PRLperdew1997}. The Kohn-Sham orbitals are expanded in a plane wave basis set with an energy cutoff of $400$\,eV. The electron-ion interactions are taken into account using the Projector Augmented Wave (PAW) method \cite{PRBblochl1994}. All geometries have been relaxed until the atomic forces were lower than $0.025$\,{eV}/{\AA}. The Brillouin Zone (BZ) is sampled according to the Monkhorst-Pack method \cite{PRBmonkorst1976}, using a gamma-centered $4 \times 4 \times 2$ mesh for atomic structure relaxation and a $5\times 5 \times 3$ mesh for obtaining the self-consistent total charge density for the bulk calculations.  The slab and grain-boundary (GB) calculations were performed with the same k-mesh, but with a single k-point in the confined directions. In the slab calculations a vacuum region of at least 12\,{\AA} was considered in order to avoid spurious interactions between periodic images. To describe the Fe $d$ orbitals' strong Coulomb interaction, we have considered the on-site repulsion term with $U=4$\,eV \cite{PRBdudarev1998}, which has shown to represent hematite magnetic phase and energy gap correctly \cite{JPCChuang2016}.

{\it Hematite surface electronic structure.} Concomitantly with the segregation of Sn/Sb atoms on hematite GB, upon sample preparations, Sn/Sb modified surfaces are present in the solid-liquid interface \cite{AEMsoares2019, JMCAnogueira2019}. Previous studies attributed this modification as responsible for the increase of $V_{\rm fb}$, Figure~\ref{fig:exp}(c). By looking at different hematite surfaces, we can extract the effect of Sn/Sb on the system Fermi level in relation to the water redox potential ($E_{\rm f, h2o}$) and thus discuss the flat band potential variation.

Keeping the \blue{stoichiometry} of the bulk we cleave the Hematite in the [001], [100] and [110] directions, and their surface energies, 
\begin{equation}
E_{s} = (E_{slab} - n\,E_{bulk})/(2A),
\end{equation} were computed; with $E_{slab}$ and $E_{bulk}$ the total energies of the slab and bulk calculations, respectively; $n$ is the slab/bulk fraction of atoms; and $A$ is the slab area. Here we find $E_{s} = 1.23$, $1.57$, and $1.14$\,$\times 10^{-4}$J/{cm$^2$}, for the (001), (100), and (110) surfaces, respectively. Therefore there is a lower cleavage energy for the (110) surface. Such preference for the (110) surface has also been experimentally verified \cite{JACSkay2006, JPCBwen2005}. As a study case we have focused in the two lowest energy surfaces (001) and (110) to consider the Sn/Sb segregation effects.

We have considered the Sn/Sb atoms to enter substitutionally at the energetic more favorable top-most surface Fe sites \cite{NLling2011}, in three possible surface terminations: stoichiometric, O-vacancy, and O-rich. The substitutional doping formation energy is negative indicating an exothermic process for the O-poor termination with $-0.53$\,eV/Sb and $-1.21$\,eV/Sn; in contrast for the surfaces with higher O concentration (stoichiometric$\rightarrow$O-rich), the formation energy become positive with $0.12 \rightarrow 0.08$\,eV/Sb and $0.48 \rightarrow 1.29$\,eV/Sn. Such doped formation energy values on the surface are lower than the calculated on the hematite bulk, $2.50$\,eV/Sb and $1.94$\,eV/Sn, which corroborates the experimentally observed surface doping.

\begin{figure}
\includegraphics[width=\columnwidth]{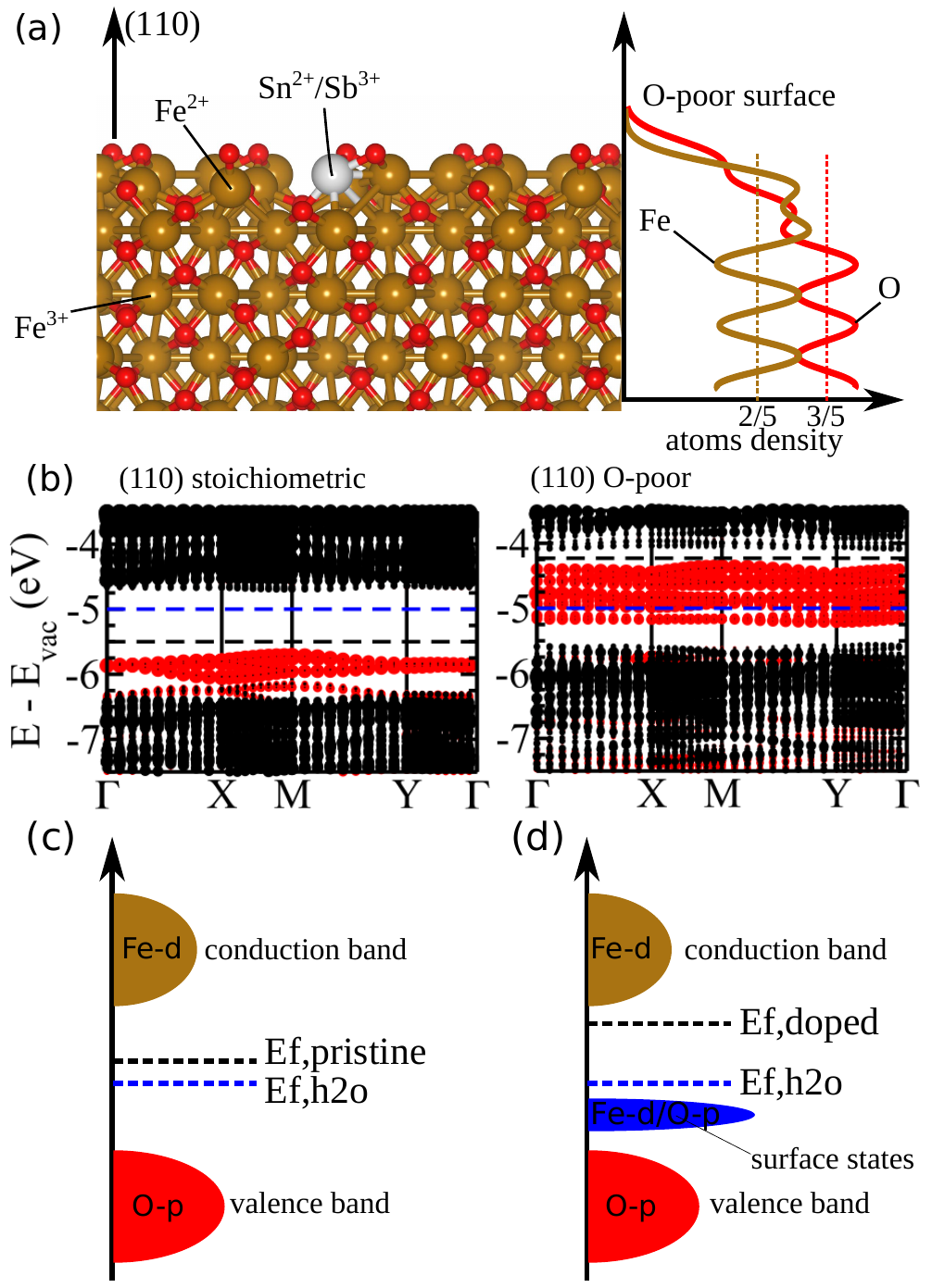}
\caption{\label{fig:surf} (a) Surface atomic configuration for the O-poor (110) termination. (b) Band structure for the pristine (left panel) and O-poor (right panel), with the projected contributions of bulk (in black) and surface (in red). (c) Pristine hematite, and (d) O-poor surface electronic schematic representation.}
\end{figure}

From the electronic point of view, in the stoichiometric case, the foreign atoms enter the Fe$^{3+}$ sites with oxidation states Sn$^{4+}$ and Sb$^{5+}$. Meanwhile, upon the presence of neighboring oxygen vacancies, the oxidation states change to Fe$^{2+}$, Sn$^{2+}$, and Sb$^{3+}$. Despite the modification of the oxidation states in the presence of oxygen vacancies, the surface incorporation of Sb/Sn has little effect on the Fermi level position with respect to the hematite bulk VBM and CBM. In Fig. \ref{fig:surf-band-001} we show the projected band structure for the (001) surface, with the slab interior (bulk states) contribution displayed in black, and the surface states contribution in red. We can see that the surface termination rules the Fermi level shift. For instance, we see the Fermi level (green dashed line) of the stoichiometric and O-rich slab termination close to the valence band for all doping cases [Fig.~\ref{fig:surf-band-001}(b) and (c)]. However, for O-poor termination, there is an upshift of the Fermi level towards the conduction band [Fig.~\ref{fig:surf-band-001}(a)]. In all studied cases, surface states (red circles) are present on the hematite bulk bandgap. Such a situation allows us to establish the picture show in Figure~\ref{fig:surf}.

\begin{figure*}
\includegraphics[width=1.9\columnwidth]{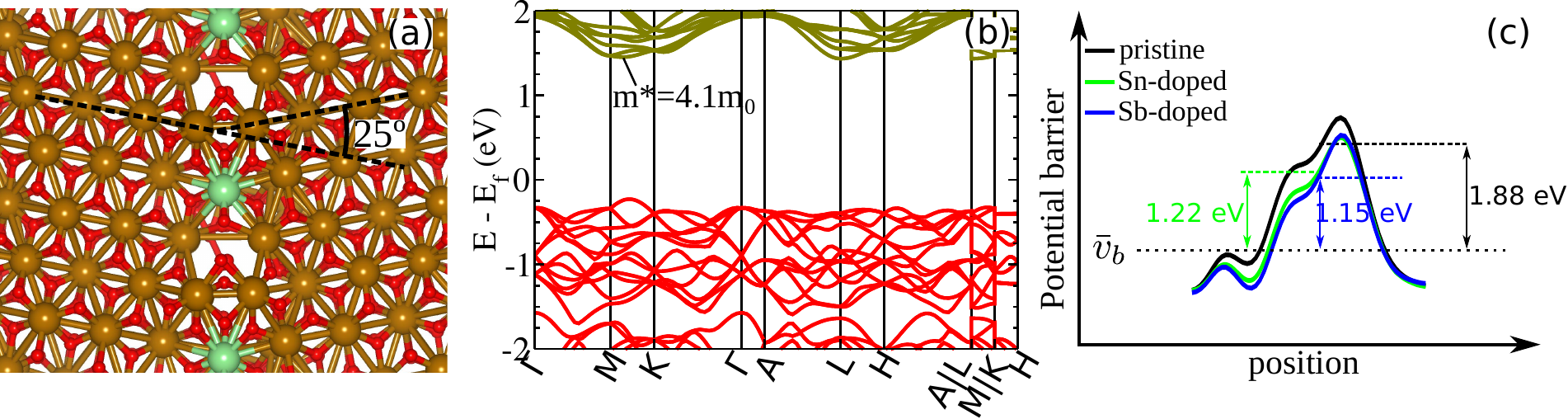}
\caption{\label{fig:gb} (a) Grain-boundaries with $25^o$ between grains. The Fe/O atoms are shown in gold/red, while the dopant site is shown in green. (b) Hematite bulk band structure with the conduction band minimum (CBM) effective mass highlighted, and the (c) effective average potential across the grain-boundary, the x-axis correspond to the horizontal direction of (a).}
\end{figure*}

In the stoichiometric terminations, the Fermi level remains in the hematite bulk bandgap. In Fig.~\ref{fig:surf}(b) left panel, we have shown the stoichiometric (110) terminated slab band structure, with the system Fermi level (black dashed line) within the valence/conduction energy gap of the bulk states (black circles). Experimentally it is known that hematite is an n-doped semiconductor \cite{CSCsivula2011}, presenting the scenario shown in  Fig.~\ref{fig:surf}(c). Upon the formation of O-vacancy on the surface [Figure~\ref{fig:surf}(a)], the $E_f$ is shifted towards the bulk conduction band [Fig.~\ref{fig:surf}(b) right panel], leading to the increase in $V_{\rm fb}$ [Figure~\ref{fig:surf}(d)]. Since the oxygen in hematite is $O^{2-}$, its vacancy act as an n-dopant. The reciprocal is true for the O-rich surface, where the Fermi level shifts towards the bulk valence band, Fig.~\ref{fig:surf-band-001}(c). We see surface states, ruled by the Fe-$d$ and O-$p$ orbitals hybridization, appearing within the bulk energy gap in all slab calculations, which can serve as hole traps \cite{ACIEwang2019}. This prediction goes according to recent experiments where NiFeOx act passivating those surface states, where the pristine character \blue{of the surface has been restored by the suppression of the hole traps} \cite{JACERSfreitas2020}.

Although Sn/Sb surface modification appears to have no direct effect on the surface Fermi energy, we see a reduction of O-vacancy formation energy by $0.77$\,eV and $0.19$\,eV, in the presence of Sn and Sb, respectively. Therefore we can establish the picture of increased O-vacancy formation, ruled by the presence of Sn in the hematite surface. Such lowering of the vacancy formation on hematite surface leads to the over-potential \cite{SSanders2015} and complete previous studies of O-vacancy facilitation on the bulk by the presence of Ti or Sn dopants \cite{RSCApan2020}.

{\it Sn/Sb doped grain boundaries.} Experiments undoubtedly predicted that hematite photocurrents undergo a transition from one grain to the other rather than percolating along the grain boundaries \cite{AEMsoares2019}. Such claims, concomitantly with the observation of increased photocurrent upon atoms' segregation at the GB, currently lack microscopic evidence. By constructing hematite grain boundaries, Fig.~\ref{fig:gb}(a), we track the electron potential barrier evolution by incorporating foreign Sn and Sb atoms at the GB \footnote{The generated GB cell considers two grains rotated by 25$^o$ with respect to each other, with two GBs in the unit cell to guarantee the periodicity, with 13.4\,{\AA} of lateral separation. Our choice were guided by experimental results, which show that hematice grain boundaries present high-angles \cite{AEMsoares2019}, with most of it higher than 15$^o$ \cite{SRwickman2017}. We have constructed the lowest possible commensurate angle which was still computationally viable, hosting 284 atoms in the unit cell.}. From Figure~\ref{fig:gb}(a) we see a lower atomic concentration at the GB, \blue{compared with the pristine region of the grain,} consistent with experimental HAADF images \cite{AEMsoares2019}. By substituting a Fe atom by Sn/Sb in the GB region [green atom in Fig.~\ref{fig:gb}(a)], i.e., keeping the same atomic concentration, we see a reduction of the effective potential energy barrier, Fig.~\ref{fig:gb}(c). Such reduction can be understood by the lower potential energy ($-Z_{\rm eff}e/r$) of the higher effective atomic number ($Z_{\rm eff}$) of Sn and Sb. We can estimate a quantitative effect of this reduction on the electron transport across the GB, approximating a rectangular barrier's potential profile. Here the barrier width is $a=2.5$\,{\AA}, and the height goes from $V_0=1.88$\,{eV} to $V_1=1.22$\,{eV} ($V_1=1.15$\,{eV}), relative to the bulk mean potential ($\bar{v}_{b}$), upon Sn/Sb segregation. In this effective free electron approximation, the wave-vector reads $k_i = \sqrt{2m^* (V_i - E)}/\hbar$, with $E$ the electron kinetic energy. We have calculated the conduction band minimum (CBM) effective mass to be $m^* = 4.1\,m_0$ [Fig.~\ref{fig:gb}(b)], close to previous studies \cite{JCPneufeld2016} and experimental estimations \cite{APbosman1970}. For a rectangular barrier, the  transmission coefficient is written as
\begin{equation}
T_i = \left( 1 + \frac{V_i^2 \sinh^2(k_i a)}{4E(V_i-E)} \right),
\end{equation}
where the $i=0$, and $1$ represents the pristine and doped GB, respectively. In this approximation we predict that the electron transmission increases upon Sn (Sb) doping by a factor of $T_1/T_0 \approx 5.5$ ($7.0$) for lower kinetic energy ($E < V_1$), and both going towards $T_1/T_0 \approx 1.5$ for higher electron kinetic energy ($E > V_0$). {The observed photocurrent can be statically viewed as $i_{ph} = \sigma (1-\eta)\,e\,v_d$, with $\sigma$ the photon absorption crossection (electron-hole pair generation), $\eta$ the electron-hole recombination rate, $e$ the electron charge and $v_d$ the drift velocity of the electrons in the material. The drift velocity is dependent on the materials conductivity $g$ and the bias between the electrodes, $v_d \propto g$. For a \blue{polycrystaline} material with non-percolating grains transport, as in hematite \cite{AEMsoares2019}, the conductivity ($g$) depends directly on the tunneling rate between grains. Therefore, the photocurrent is proportional to the transmission rate of the electrons between grains.} Astonishingly, the experimental increase of the photocurrent occurs by a factor of $2$ for Sn doped \cite{JACERSfreitas2020} and $6.5$ for Sb doped \cite{JMCAnogueira2019}, at the same order of the values predicted by our approximated model. Such microscopic interpretation allows us to understand and predict quite reasonably the experimentally observed increase of the photocurrent upon Sn/Sb doping.

{\it In summary,} we have presented a clear picture of the dopants segregation effect on the solid-solid and solid-liquid interfaces of hematite. We show that in hematite's grain boundaries (solid-solid interface), the Sn/Sb dopants reduce the electronic barrier, increasing the inter-grain electron transmission. Our results predict that the segregation of other dopants with higher atomic numbers will also contribute to the photocurrent increase. On the other hand, the dopants' effect on the hematite's surface (solid-liquid interface) electronic is indirect. The Sn/Sb dopants lower the oxygen vacancy formation, leading to the flat band potential increase. Our results describing the electronic effects of segregation can guide the rational functionalization of hematite towards higher-efficiency photoanodes.

\section*{ACKNOWLEDGMENTS}
The authors acknowledge financial support from FAPESP (projects CEPID - 2013/07296-2, 2017/03135, 17/18139-6, 17/02317-2, 2018/05159-1 and 2019/20857-0), FINEP, CNPq (project 401452/2017-4, process 168295/2017-2), and CAPES (grand numbers 88887.310931/2018-00 and 88882.332718/2019–01). The authors also acknowledge the Laborat\'{o}rio Nacional de Computa\c{c}\~{a}o Cient\'{i}fica (project SCAFMat2) for computer time.


\section*{Data Availability Statement}
The data that support the findings of this study are available from the corresponding author upon reasonable request

\bibliography{bib}

\end{document}